\documentclass[runningheads]{llncs}
\usepackage{graphicx}
\usepackage{hyperref}
\graphicspath{{./fig/}}

\begin{document}
\title{Towards the Tunka-Rex Virtual Observatory}
%
%
\author{
P.~Bezyazeekov\inst{1}
\and
N.~Budnev\inst{1}
\and
O.~Fedorov\inst{1}
\and
O.~Gress\inst{1}
\and
O.~Grishin\inst{1}
\and
A.~Haungs\inst{2}
\and
T.~Huege\inst{2,3}
\and
Y.~Kazarina\inst{1}
\and
M.~Kleifges\inst{4}
\and
D.~Kostunin\inst{5}
\and
E.~Korosteleva\inst{6}
\and
L.~Kuzmichev\inst{6}
\and
V.~Lenok\inst{2}
\and
N.~Lubsandorzhiev\inst{6}
\and
S.~Malakhov\inst{1}
\and
T.~Marshalkina\inst{1}
\and
R.~Monkhoev\inst{1}
\and
E.~Osipova\inst{6}
\and
A.~Pakhorukov\inst{1}
\and
L.~Pankov\inst{1}
\and
V.~Prosin\inst{6}
\and
F.~G.~Schr\"oder\inst{2,7}
\and
D.~Shipilov\inst{1}
\and
A.~Zagorodnikov\inst{1}
}%
\authorrunning{P. Bezyazeekov et al.}
%
\institute{
Institute of Applied Physics ISU, Irkutsk, Russia
\and 
KIT, Institut f\"ur Kernphysik, Karlsruhe, Germany
\and 
Astrophysical Institute, Vrije Universiteit Brussel, Pleinlaan 2, Brussels, Belgium
\and
Institut f\"ur Prozessdatenverarbeitung und Elektronik, KIT, Karlsruhe, Germany
\and
DESY, Zeuthen, Germany
\and
Skobeltsyn Institute of Nuclear Physics MSU, Moscow, Russia
\and
Bartol Research Inst., Dept. of Phys. and Astron., Univ. of Delaware, Newark, USA
}
\maketitle              
\begin{abstract}
The Tunka Radio Extension (Tunka-Rex) is a cosmic-ray detector operating since 2012.
The detection principle of Tunka-Rex is based on the radio technique, which impacts data acquisition and storage.
In this paper we give a first detailed overview of the concept of the Tunka-Rex Virtual Observatory (TRVO), a framework for open access to the Tunka-Rex data,
which currently is under active development and testing.
We describe the structure of the data, main features of the interface and possible applications of the TRVO.

\end{abstract}
\section{Introduction}

Following the approach chosen in the German-Russian Astroparticle Data Life Cycle initiative (GRADLCI)~\cite{Bychkov:2018zre} 
we are preparing to publish the data of the Tunka Radio Extension (Tunka-Rex) experiment under a free data license.

\begin{figure}[t!]
\centering
\includegraphics[width=0.95\linewidth]{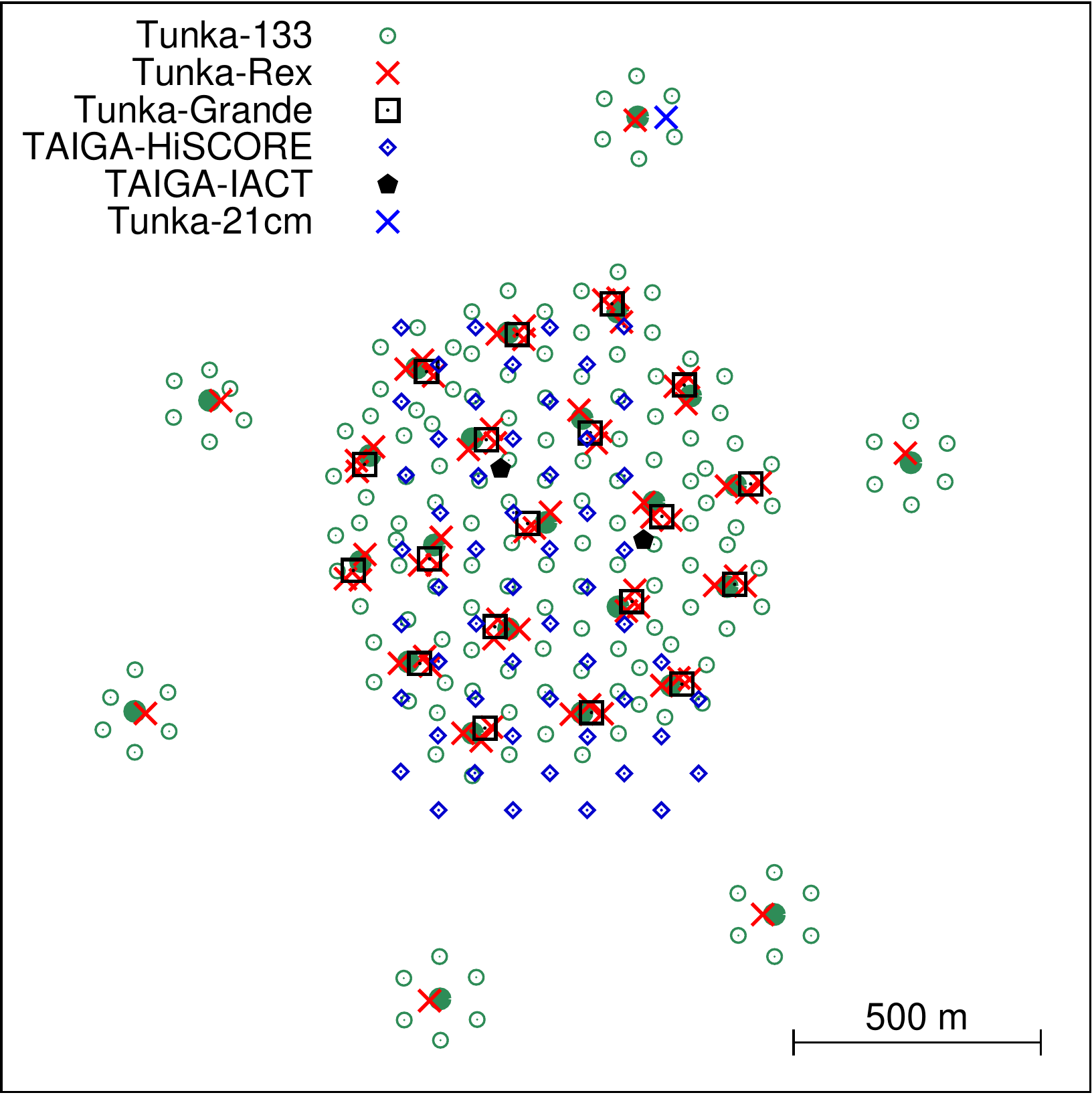}\\
\includegraphics[width=0.95\linewidth]{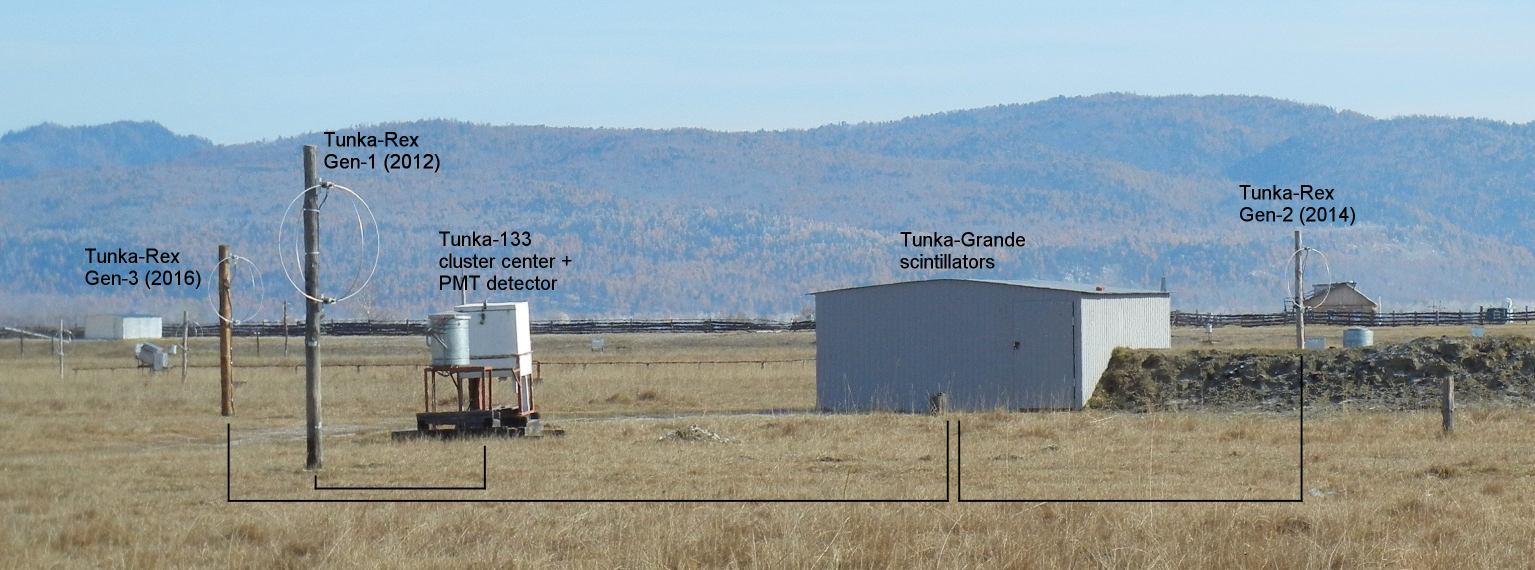}
\caption{
The layout of the TAIGA setups as of 2019.
The antenna stations are depicted as crosses (the Tunka-21cm array is depicted by a single marker due to scale of the map).
The layout and hardware configuration was changes several times during the past years.
\textit{Bottom:} Photo of a single cosmic-ray cluster of the TAIGA facility. 
Lines mark the cable connections.
}
\label{fig:tunka}
\end{figure}

Tunka-Rex is a digital antenna array located at the Tunka Advanced Instrument for cosmic rays and Gamma Astronomy (TAIGA) observatory~\cite{Budnev:2017fyg,Kostunin:2019nzy}.
The TAIGA setups can be divided in two main classes of installations: 
dedicated to cosmic rays (Tunka-133~\cite{Prosin:2015voa}, Tunka-Rex~\cite{Bezyazeekov:2015rpa} and Tunka-Grande~\cite{Budnev:2015cha}) 
and dedicated to gamma rays (Tunka-HiSCORE~\cite{Tluczykont:2017pin} and TAIGA-IACT~\cite{Yashin:2015lzw}).
In Fig.~\ref{fig:tunka} one can see the layout of the facility and note that the cosmic-ray setups are grouped in clusters: 19 clusters in a dense core and 6 satellite clusters.
Each core cluster is equipped with 3 Tunka-Rex antenna stations, while satellite clusters contain one antenna station, each, and no Tunka-Grande scintillators.

\begin{figure}[t!]
\includegraphics[width=1.0\linewidth]{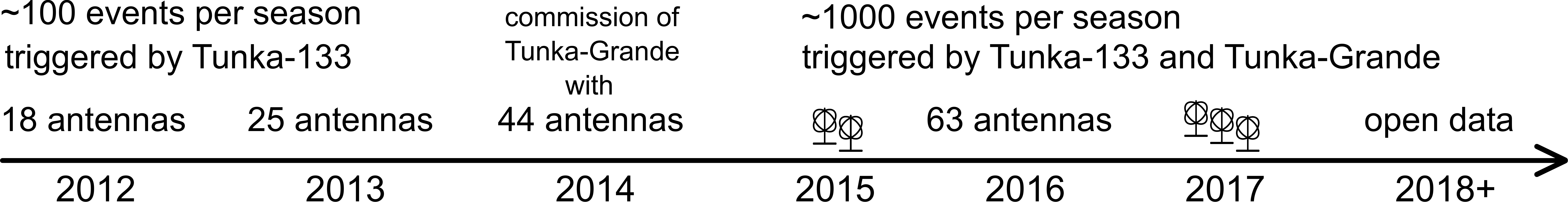}
\caption{The antenna array has been commissioned in 2012 with 18 antenna stations triggered by the Tunka-133 air-Cherenkov detectors.
Since the commissioning of Tunka-Grande in 2014-2015, Tunka-Rex additionally receives a trigger from Tunka-Grande (during daytime measurements).
Starting from 2018, we are working on the public access of Tunka-Rex software and data.}
\label{fig:timeline}
\end{figure}

For the time being, Tunka-Rex consists of 57 antenna stations located in the dense core of TAIGA (1~km\textsuperscript{2}) and 6 satellite antenna stations expanding the sensitive area of the array to 3~km\textsuperscript{2}.
Tunka-Rex has been commissioned in 2012 with 18 antenna stations triggered by the air-Cherenkov array Tunka-133.
In the following years, Tunka-Rex was upgraded several times. 
The TAIGA facility was enhanced by the Tunka-Grande scintillator array providing a trigger for Tunka-Rex since 2015.
One can see the timeline of the Tunka-Rex development in Fig.~\ref{fig:timeline}.

\begin{figure}[t!]
\includegraphics[width=1.0\linewidth]{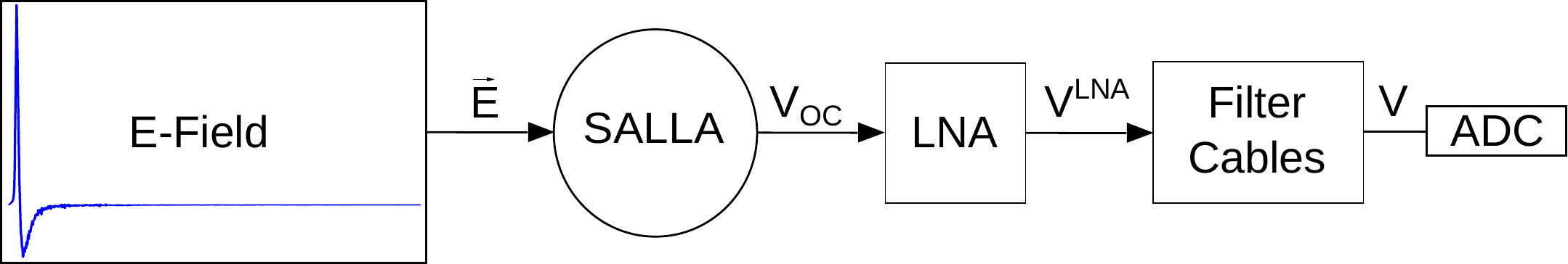}
\caption{Scheme depicting the instrument response function (IRF).
The incoming radio signal is received by the antenna, passes through electronics and cables, and is recorded by the ADC.
For details see Ref.~\cite{Bezyazeekov:2015rpa}.
}
\label{fig:irf}
\end{figure}

Each Tunka-Rex antenna station consists of two perpendicular active Short Aperiodic Loaded Loop Antennas (SALLA)~\cite{Abreu:2012pi} pre-amplified with a Low Noise Amplifier (LNA).
Signals from the antenna arcs are transmitted via 30~m coaxial cables to an analog filter-amplifier, which cuts the frequency band to 30-80~MHz.
The filtered signal is then digitized by the local data acquisition system (DAQ) with a 12 bit-sampling at a rate of 200 MHz; the data are collected in traces of 1024 samples each.
Each element of this signal chain has been calibrated under laboratory conditions, which resulted in the instrument response function (IRF) defining the resulting digital traces recorded by the DAQ (see Fig.~\ref{fig:irf})
For the reconstruction of the original signal, the inverse IRF is convoluted with the raw data.
This convolution defines the \textit{data layers (DL)} defined below.

The distinguishing feature of the broadband radio detectors is that they can be used both for radio astronomy and astroparticle purposes (e.g. ultra-high neutrino and cosmic-ray detection) depending on the configuration and operation mode.
For example, the core of the LOFAR antenna array has been successfully applied for cosmic-ray detection~\cite{Schellart:2013bba}; meanwhile the proposed air-shower array GRAND aims also at astronomy goals~\cite{Alvarez-Muniz:2018bhp}
Therefore, we will extend the concept of KCDC~\cite{Haungs:2018xpw} and implement additional features in our framework for open data, which will result in the Tunka-Rex Virtual Observatory (TRVO).

\section{Structure of the Tunka-Rex data}
\label{sec:data_structure}
In this section we provide a general description of the Tunka-Rex data types, their structure, and their connection with the hardware of the experiment and observed phenomena.
\subsection{Antenna station data}
As described above, raw Tunka-Rex data consist of traces recorded for each antenna from the DAQ buffer after receiving an external trigger.
The data on an antenna station can be described by the following fields:
\begin{itemize}
\item \texttt{Trace ID}: unique identifier of the trace
\item \texttt{Antenna ID}: identifier of the antenna station, enumerated with the following convention: 1-25 (1st generation), 31-49 (2nd generation), 61-79 (3rd generation)
\item \texttt{Timestamp}: float number of the GPS time of the event with nanosecond precision
\item \texttt{Version}: the version of the \textit{data release (DR)}
\item \texttt{Traces}: serialized arrays (two channels or three electric-field components) each with 1024 elements, either integer of float number depending on the \texttt{DL}
\item \texttt{Flags}: additional flags describing the status of the antenna station, e.g. operation, malfunction, calibration, etc.
\end{itemize}
As will be described below, \texttt{DL0-2} differs only in the way of the representation of the \texttt{Traces} field.

\subsection{Calibration data}
The calibration data defines the instrument response function and is used for simulation and for reconstruction.
Moreover, it reflects the location of the antenna station (antennas can be re-located and re-aligned) and its hardware configuration, since some components were occasionally replaced due to malfunction.
Thus, each antenna station is described by the following calibration data:
\begin{itemize}
\item \texttt{Commission}: timestamp of the commission of configuration
\item \texttt{Decommission}: timestamp of the decommission of configuration
\item \texttt{Antenna ID}: identifier of the antenna station (identical to ID in raw data)
\item \texttt{LNA ID}: identifier of the low noise amplifier
\item \texttt{Filter ID}: identifier of the filter-amplifier
\item \texttt{X, Y, Z}: coordinates of the antenna station in local coordinates
\item \texttt{Alignment}: alignment of the antenna station with respect to the magnetic North (the initial alignment of $45^\circ$ slightly changed over time)
\end{itemize}
Besides these time-dependent properties of the antenna station, the calibration is defined by the phase and gain response of the antennas and the signal chain.

\subsection{Supplementary data}
The supplementary data describe observation conditions, and are shared with the other TAIGA setups.
A detailed description of this type of data is given in the same proceedings in Ref.~\cite{Shigarov_DLC2019}.
The most important supplementary data for Tunka-Rex are \texttt{Trigger} (operation mode, thresholds, online/offline clusters, etc.) and \texttt{Environment} (temperature, pressure, humidity, magnetic field, etc).

\subsection{Air-shower data}
Since Tunka-133 and Tunka-Grande, which provide the trigger for Tunka-Rex, feature an independent reconstruction of air-shower events, the combination of the data from all three setups can improve the reconstruction of the primary cosmic ray.
The data structure for the particle detectors is described and implemented in the frame of KCDC, and the Tunka-Grande event reconstruction perfectly fits to this system.
Because Tunka-133 and Tunka-Rex perform calorimetric measurements, their fields differ and are described as:
\begin{itemize}
\item \texttt{UUID}: universally unique identifier\footnote[1]{\url{https://www.itu.int/en/ITU-T/asn1/Pages/UUID/uuids.aspx}} of the event.
The UUID is chosen in order to avoid collisions during distributed data acquisition
\item \texttt{Timestamp}: float number of the GPS time of the event with nanosecond precision
\item \texttt{Theta, Phi}: Arrival direction (zenith and azimuth angles)
\item \texttt{X, Y, Z}: Coordinates of the shower core
\item \texttt{Energy}: Energy of the primary particle
\item \texttt{Xmax}: Depth of the shower maximum
\item \texttt{Particle}: Type of the primary particle
\end{itemize}
Besides the reconstruction of the air-shower and primary particle properties the signals at the individual antenna stations of Tunka-Rex and at the optical modules of Tunka-133 are described by the following fields:
\texttt{Timestamp}, \texttt{Amplitude}, \texttt{SNR}, \texttt{Width}, \texttt{Power}, etc.

\section{Data layers}
In this section we describe the naming conventions for the data layers in the TRVO.
\texttt{DL0-2} are organized in the standard structure described above: Station $+$ Calibration $+$ Supplementary data,
while the \texttt{DL3+} can have additional entries, e.g. cosmic-ray events, radio bursts, etc.

\textbf{Data Layer 0} consists of raw traces recorded by the ADCs, i.e. arrays containing values in the range [0;4095].
These data are intended to be used in case of recalibration/debugging of the instrument and are not recommended for the external application.

\textbf{Data Layer 1} consists of the traces containing voltages at the antenna stations (i.e. antenna-induced voltages) obtained after unfolding the raw traces from the hardware response of Tunka-Rex amplifiers, filters, and cables.
From these values the electrical field at the antenna station can be reconstructed using the specific antenna pattern and direction of incoming radio wave.

\textbf{Data Layer 2} consists of the traces containing voltages converted to the values of electrical field at the antenna stations.
Depending on the data release, the electrical fields will be calculated for air-shower events (\texttt{DL2-AIRSHOWER}), for astronomical objects (\texttt{DL2-ASTRONOMY}), or for any other kind of measurements, e.g. background, RFI, etc (\texttt{DL2-OTHER}).

\textbf{Data Layer 3+} will contain high-level reconstruction of radio data, i.e. quantities obtained after sophisticated processing and analyzing of radio traces.
These data can be represented in tables, histograms, FITS files, etc.

\section{Storage of the data}
Since the main Tunka-Rex data are represented as a linear set, we have decided to use a relational database based on an open engine such as \texttt{MySQL} or \texttt{PostgreSQL}.
The raw data from the single antenna station have a relatively small size (few KiB) and can be stored entirely in a single row of the SQL table.
We have deployed several testing databases with Tunka-Rex events on the servers of the Irkutsk State University (ISU) and the Karlsruhe Institute of Technology (KIT).
The expected number of entries in the database from several data releases is in the order of billions which result in TiB scale of DB.
Currently we are testing the performance of the database and implementing a user interface and basic features.

\section{Access to the data}
As mentioned in the previous section, for the time being we are working on the implementation of a client for TRVO, which features basic access to the primary data and a plugin extension for more sophisticated quality cuts.
Plugins will provide an interface to the DB and allow for end-user implemented scripts for online data analysis, quality cuts, and other preprocessing manipulations of data.
Below we give the description of two initial plugins which will be delivered by default.

\subsection{Cosmic-ray event builder}
Since the metadata of cosmic-ray events reconstructed by Tunka-Rex will be integrated in the common GRADLCI framework,
TRVO will only provide an index of events reconstructed by Tunka-Rex (\texttt{DL3}) and the connection between corresponding data layers.
The query engine supports backward compatibility, and data can either be selected by TRVO directly of via the GRADLCI metadata engine (with support of joint analysis including third-party data).

\subsection{Radio astronomy tools}
Besides access to cosmic-ray events, we will provide astronomy-related tools for the direct manipulation with radio traces: band-stop, band-pass, and median filters, beam-former, skymap builder, and others.

\subsection{Software and datasets published already}
The previously published Tunka-Rex datasets and software can be found at the following URL: \url{http://soft.tunkarex.info}; the official Mercurial repository of the Tunka-Rex software can be found on Bitbucket: \url{https://bitbucket.org/tunka}.
We plan to use the \texttt{astroparticle.online} platform for future releases.

\section{Application of the Tunka-Rex Virtual Observatory}
Since the primary goal of Tunka-Rex is the detection of cosmic rays,
the main application of TRVO is providing access to the high-level reconstruction of air showers (\texttt{DL3+}).
The architecture of this part of the Virtual Observatory has already been developed in the frame of KCDC and we do not plan to depart from this concept significantly.
Besides public access to cosmic-ray data of the TAIGA observatory, the radio data can be used for cross-calibration of different cosmic-ray experiments, as shown in Ref.~\cite{Apel:2016gws}.
Below we discuss unique features of the Tunka-Rex archival data and their application to current and future research (it is worth noting, that the Tunka-Rex trigger is tuned for cosmic-ray detection and the selection from the archival data might be significantly biased and can be used only for tentative studies).
\begin{itemize}
\item \textit{Studies of the radio background in the frequency band of 30-80~MHz.}
Nowadays there are only few radio telescopes operating in this frequency band, moreover these telescopes operate in an interferometric mode. 
They correlate the radio signal using beam-forming and record the resulting correlation, while radio arrays aimed at cosmic-ray detection record full uncorrelated time series.
The broadband measurement of radio background in this frequency band is of special interest to search for a possible cosmological signal from neutral hydrogen.
Since this signal has a signal-to-noise ratio (SNR) of about $10^{-5}$, understanding of systematic uncertainties is crucial for this type of measurements.
The Tunka-Rex child experiment, Tunka-21cm, tests the possibility of application of cosmic-ray detectors for studies of this cosmological signal, and is a first user of \texttt{DL2-BACKGROUND} and \texttt{DL2-ASTRONOMY}.
\item \textit{Searching for radio transients.}
Obviously archival data can be used for searching for astronomical transients in this frequency band.
The effective exposure of Tunka-Rex provides only a very small probability of detection of any kind of transients.
However, the archival data can be used for the test of detection techniques for future multi-purpose detectors.
\item \textit{Training of neural networks for RFI tagging.}
It was shown, that deep learning can improve the signal reconstruction of radio detectors when using an autoencoder architecture~\cite{Shipilov:2018wph,Erdmann:2019nie,Bezyazeekov_DLC2019},
because neural networks are able to learn features of the background and can be used for either denoising of radio traces or tagging of traces containing special features.
It is worth noting, that the present Tunka-Rex autoencoder is trained on a dataset containing less than 1\% of all Tunka-Rex background traces, what promises significant improvements by using larger training samples extracted from TRVO.
\item \textit{Outreach and education.} 
Open data implies outreach and educational activities, and we support this activities. 
TRVO will be used as educational platform in the outreach part of the GRADLCI~\cite{Kazarina_DLC2019} and \texttt{astroparticle.online} projects.
At the first stage we use the Tunka-Rex hardware, software, and simulations for the training of students of the Physics Department of ISU.
\end{itemize}
Last but not least, the developed framework can be applied to future arrays: GRAND~\cite{Alvarez-Muniz:2018bhp} and radio extensions of the Pierre Auger Observatory~\cite{Hoerandel_ARENA2018} and the Tien-Shan cosmic-ray setup~\cite{Beisenova:2017knp}.
\section{Conclusion}
The Tunka-Rex Virtual Observatory provides open access to the data of experiments measuring cosmic rays with radio technique.
We plan to combine both astroparticle- and astronomy-related features in TRVO and provide fast and user-friendly access with the possibility of custom scripting for complex preselection and preprocessing of the data.
The first databases have already been deployed and are now under internal testing.
Besides users from the education sector (ISU) and partner experiments (TAIGA) we have requests from the recently established engineering setup Tunka-21cm aimed at astronomical goals.

\section*{Acknowledgements}
This work was supported by Russian Science Foundation Grant 18-41-06003 
(Section \ref{sec:data_structure}), Helmholtz Society Grant HRSF-0027 and
by Russian Foundation for Basic Research Grant 18-32-20220.
We thank the members of KCDC and GRADLCI for the fruitful discussions and support of the deployment of testing databases.
%
%
%
\bibliographystyle{ieeetr}
\bibliography{references}
\end{document}